\documentclass{article}
\usepackage{amsmath, amssymb, amsfonts}
\title{Pattern for a star filled with imperfect fluid }
\author{Hristu Culetu, \\Ovidius University, Dept.of Physics and Electronics, \\B-dul Mamaia 124, 900527 Constanta, Romania, \\e-mail : hculetu@yahoo.com}

\begin{document}
\numberwithin{equation}{section}
\pagenumbering{arabic}
\maketitle
\newcommand{\fv}{\boldsymbol{f}}
\newcommand{\tv}{\boldsymbol{t}}
\newcommand{\gv}{\boldsymbol{g}}
\newcommand{\OV}{\boldsymbol{O}}
\newcommand{\wv}{\boldsymbol{w}}
\newcommand{\WV}{\boldsymbol{W}}
\newcommand{\NV}{\boldsymbol{N}}
\newcommand{\hv}{\boldsymbol{h}}
\newcommand{\yv}{\boldsymbol{y}}
\newcommand{\RE}{\textrm{Re}}
\newcommand{\IM}{\textrm{Im}}
\newcommand{\rot}{\textrm{rot}}
\newcommand{\dv}{\boldsymbol{d}}
\newcommand{\grad}{\textrm{grad}}
\newcommand{\Tr}{\textrm{Tr}}
\newcommand{\ua}{\uparrow}
\newcommand{\da}{\downarrow}
\newcommand{\ct}{\textrm{const}}
\newcommand{\xv}{\boldsymbol{x}}
\newcommand{\mv}{\boldsymbol{m}}
\newcommand{\rv}{\boldsymbol{r}}
\newcommand{\kv}{\boldsymbol{k}}
\newcommand{\VE}{\boldsymbol{V}}
\newcommand{\sv}{\boldsymbol{s}}
\newcommand{\RV}{\boldsymbol{R}}
\newcommand{\pv}{\boldsymbol{p}}
\newcommand{\PV}{\boldsymbol{P}}
\newcommand{\EV}{\boldsymbol{E}}
\newcommand{\DV}{\boldsymbol{D}}
\newcommand{\BV}{\boldsymbol{B}}
\newcommand{\HV}{\boldsymbol{H}}
\newcommand{\MV}{\boldsymbol{M}}
\newcommand{\be}{\begin{equation}}
\newcommand{\ee}{\end{equation}}
\newcommand{\ba}{\begin{eqnarray}}
\newcommand{\ea}{\end{eqnarray}}
\newcommand{\bq}{\begin{eqnarray*}}
\newcommand{\eq}{\end{eqnarray*}}
\newcommand{\pa}{\partial}
\newcommand{\f}{\frac}
\newcommand{\FV}{\boldsymbol{F}}
\newcommand{\ve}{\boldsymbol{v}}
\newcommand{\AV}{\boldsymbol{A}}
\newcommand{\jv}{\boldsymbol{j}}
\newcommand{\LV}{\boldsymbol{L}}
\newcommand{\SV}{\boldsymbol{S}}
\newcommand{\av}{\boldsymbol{a}}
\newcommand{\qv}{\boldsymbol{q}}
\newcommand{\QV}{\boldsymbol{Q}}
\newcommand{\ev}{\boldsymbol{e}}
\newcommand{\uv}{\boldsymbol{u}}
\newcommand{\KV}{\boldsymbol{K}}
\newcommand{\ro}{\boldsymbol{\rho}}
\newcommand{\si}{\boldsymbol{\sigma}}
\newcommand{\thv}{\boldsymbol{\theta}}
\newcommand{\bv}{\boldsymbol{b}}
\newcommand{\JV}{\boldsymbol{J}}
\newcommand{\nv}{\boldsymbol{n}}
\newcommand{\lv}{\boldsymbol{l}}
\newcommand{\om}{\boldsymbol{\omega}}
\newcommand{\Om}{\boldsymbol{\Omega}}
\newcommand{\Piv}{\boldsymbol{\Pi}}
\newcommand{\UV}{\boldsymbol{U}}
\newcommand{\iv}{\boldsymbol{i}}
\newcommand{\nuv}{\boldsymbol{\nu}}
\newcommand{\muv}{\boldsymbol{\mu}}
\newcommand{\lm}{\boldsymbol{\lambda}}
\newcommand{\Lm}{\boldsymbol{\Lambda}}
\newcommand{\opsi}{\overline{\psi}}
\renewcommand{\tan}{\textrm{tg}}
\renewcommand{\cot}{\textrm{ctg}}
\renewcommand{\sinh}{\textrm{sh}}
\renewcommand{\cosh}{\textrm{ch}}
\renewcommand{\tanh}{\textrm{th}}
\renewcommand{\coth}{\textrm{cth}}

\begin{abstract}
 A static, spherically symmetric spacetime with negative pressures is conjectured inside a star. The gravitational field is repulsive and so a central singularity is avoided. The positive energy density and the pressures of the imperfect fluid are finite everywhere. The Tolman-Komar energy of the space is negative, as for a de Sitter geometry. From the Darmois-Israel junction conditions on the star surface one finds the constant length $b$ from the metric and the expression of the surface tension $\sigma$ of the thin shell separating the interior from the Schwarzschild exterior. Some properties of the timelike and null geodesics in the Painleve-Gullstrand coordinates are investigated.

\end{abstract}

 \section{Introduction}
   Current data supports the view that the matter content of the Universe consists of two basic components, namely dark matter (DM) and dark energy (DE) with ordinary matter playing a minor role. The nature and composition of DM and DE is still not understood. The entire motivation for the existence of DM and DE is based on their validity at all distance scales of the standard Newton - Einstein gravitational theory \cite{PM} and on the experimental fact that our Universe is accelerating.
	Mazur and Mottola \cite{MM1} extended the Bose-Einstein condensation to a gravitational system and constructed a dark, compact object, with a de Sitter interior and an exterior Schwarzschild geometry, being separated by a thin shell \cite{MM2, EM}. More recently, Danila et al. \cite{BD} consider the physical properties of some classes of neutron and Bose-Einstein condensate stars in the hybrid metric-Palatini gravitational theory, which is very successful in observed phenomenology, unifying local constraints at the Solar system level and the late time cosmic acceleration. 
	
	Brandenberger and Frohlich \cite{BF} investigated the possibility that DM and DE would have a common origin. They introduced a complex axion field whose radial component gives rise to DE while the angular component plays the role of DM. Lemos and Zaslavskii \cite{LZ} studied a general relativistic solution of gravitational equations, composed of a Zel'dovich-Letelier star interior with a static configuration coupled to a Schwarzschild exterior through a spherical thin shell. The interior solution encloses some matter in a spacetime pit, dubbed pit solution. This string pits resemble Wheeler's bags of gold. Comer and Katz \cite{CK} stated that there is no reason to suppose that the energy-momentum tensor always takes the perfect fluid form inside a star, admiting anisotropic pressures (when radial and tangential pressures are different), with some of them being even negative. A system with such features might be applied to relativistic regime (for example, for neutron stars).
	
	Motivated by the previous studies on the interior properties of a relativistic star, we propose in Sec.2 the spherically-symmetric inner space is composed from an imperfect fluid with negative pressures but positive energy density. The geometry is regular throughout the star interior, being flat at the origin $r = 0$ and presenting a horizon at infinity. The classical field $\Phi$ we introduced is repulsive, having transversal pressures as the sources on the r.h.s. of the Poisson equation. In Sec.3 we apply the above model to a star interior, paying attention to the Darmois-Israel junction conditions at a thin shell separating the inner metric from the outer Schwarzschild geometry and focus especially on the shell properties - the surface tension and its energy density. Sec.4 is devoted to the timelike and null geodesics for the spacetime inside the star, with the metric written in Painleve-Gullstrand (PG) coordinates. We end with few comments in Sec.5. Geometrical units $G = c = 1$ are used throughout the paper, unless otherwise specified.
	
	\section{Regular imperfect fluid}
We investigate a spacetime of the following form
	 \begin{equation}
  ds^{2} = -(1- e^{-\frac{b}{r}}) ~dt^{2} + \frac {dr^{2}}{1- e^{-\frac{b}{r}}} + r^{2} d \Omega^{2}
 \label{2.1}
 \end{equation}
where $b$ is a positive constant and $d \Omega^{2}$ stands for the metric on the unit 2 - sphere. The coordinate ($t, r, \theta, \phi$) have the standard meaning. As we shell see, the term $e^{-\frac{b}{r}}$ plays the role of a regulator \cite{HC1}. We have, indeed, $f(r) \rightarrow 1$ if $r \rightarrow 0$ and $f(r) \rightarrow 0$ at infinity, where $f(r) = -g_{tt} = 1- e^{-\frac{b}{r}}$. The function $f(r)$ has an inflexion point at $r = b/2$, with $0<f(r)<1$. The line-element (2.1) is Minkowskian at the origin and has a horizon at infinity. 

Let us consider a static observer in the geometry (2.1) with a velocity vector field
 \begin{equation}
 u^{a} = \left(\frac{1}{\sqrt{1- e^{-\frac{b}{r}}}}, 0, 0, 0\right)
\label{2.2}
\end{equation}
which gives us the corresponding covarian acceleration
 \begin{equation}
 a^{b} = \left(0, -\frac{b}{2r^{2}} e^{-\frac{b}{r}}, 0, 0\right),
\label{2.3}
\end{equation}
with the only nonzero component $a^{r} = f'(r)/2$, where $f' \equiv df/dr$. One also notes that $a^{r}<0$, that means our observer should accelerate towards the origin for to preserve the same position. In other words, the gravitational field is repulsive. The same effect undergoes a static observer in de Sitter geometry, in static coordinates. The function $a^{r}(r)$ vanishes at $r = 0$ and at infinity. In addition, it takes a minimum value $a^{r}_{min} = -4/e^{2}b$ (lne = 1), at $r = b/2$. In the region $r>>b$ one obtains $f(r)\approx b/r$ (first order in $b/r$), a situation analysed by Vaz \cite{CV} who identified a process by which an energy extraction from the center occurs, leaving behind a negative point mass at the center to which corresponds an energy density $\epsilon (r) \propto 1/r^{2}$ and a radial pressure $p_{r} = -\epsilon $, with zero transversal pressures (see also \cite{HC2, DNY, HC3}).

For the geometry (2.1) the scalar curvature is given by
 \begin{equation}
 R^{b}_{~b} = \frac{1}{r^{2}}\left[1 + \left(1 + \frac{b}{r}\right)^{2}\right] e^{-\frac{b}{r}},
\label{2.4}
\end{equation}
that is finite in the whole space (same is valid for the Kretschmann scalar). Moreover, (2.2) leads to a vanishing shear tensor $\sigma^{a}_{b}$. If we wrote formally $f(r)$ as $f(r) = 1 - \frac{2m(r)}{r}$, where $m(r)$ is the mass up to the radius $r$ (or the Misner-Sharp mass in our case), we would get $m(r) = \frac{r}{2} e^{-\frac{b}{r}}$, which shows clearly that, when $r \rightarrow \infty$ (or, more physically, if $r>>b$), we have $r \approx 2m(r)$, which is the horizon location. As a function of $r$, $m(r) \rightarrow 0$ when $r \rightarrow 0$ and, at infinity, $m(r)$ approaches its asymptote $m(r) = \frac{r}{2} - \frac{b}{2}$.

Let us find now the source of the metric (2.1), namely the stress tensor we need on the r.h.s. of Einstein's equation $G_{ab} = 8\pi T_{ab}$ in order to have (2.1) its exact solution. We write down firstly the general expression of the energy-momentum tensor for an imperfect fluid \cite{KM, HC4, FC}
\begin{equation}
T_{~a}^{b} = (\rho + p_{t})u_{a} u^{b} + p_{t} \delta_{a}^{b}+ (p_{r} - p_{t}) n_{a} n^{b} + u_{a}q^{b} + u^{b}q_{a},
\label{2.5}
\end{equation}
	where $\rho = T_{ab}u^{a}u^{b}$ is the energy density, $p_{t}$ represents the transversal pressures ($p_{t} = p_{\theta} = p_{\phi}$), $n^{b}$ is a spacelike vector with $ n_{a} n^{a} = 1$ and $n_{a} u^{a} = 0$, $q_{a}$ is the energy flux density four vector, $p_{r}$ is the radial pressure and $u^{b}$ is given by (2.2). From its properties we get $n^{a} = (0, \sqrt{1 - e^{-\frac{b}{r}}}, 0, 0)$. On the grounds of the above informations, one finds that
	\begin{equation}
	\rho = \frac{1}{8\pi r^{2}}\left(1 + \frac{b}{r}\right) e^{-\frac{b}{r}} = -p_{r},~~~p_{t} = -\frac{b^{2}}{8\pi r^{4}}e^{-\frac{b}{r}},~~~q^{a} = 0.
\label{2.6}
\end{equation}
It is worth noting that $\rho$ is always positive but all pressures are negative. In the region $r>>b$ (or $b/r \rightarrow 0$) we have $\rho = -p_{r} \approx 1/8\pi r^{2}$ and $p_{\theta} \approx 0$, values obtained previously in \cite{CV, HC2, DNY}. Moreover, $\rho$ and $p_{\theta}$ are finite everywhere, being zero at the origin and at infinity. The function $\rho (r)$ acquires a maximum value $\rho_{max} \approx 1.8/8\pi b^{2}$ at $r_{max} = b(\sqrt{3} - 1)/2$. In addition, $d\rho/dr>0$ before $\rho$ reaches its maximum value, due to the repulsive forces inside the star. That prevents the formation of a singularity at the center, where actually the geometry is Minkowskian.

We look now for a Newtonian potential given by $f(r) = 1 + 2\Phi$, whence $\Phi = -(1/2)e^{-\frac{b}{r}}$. We may write the Poisson equation for $\Phi$
	\begin{equation}
	\nabla^{2}\Phi = \frac{1}{r^{2}}\frac{d}{dr}(r^{2}\Phi) = 4\pi \cdot 2p_{\theta},
\label{2.7}
\end{equation}
	with $d\Phi /dr = -|a^{r}|$. Noting that the sources from the r.h.s. of (2.7) are the transversal pressures but not the energy density $\rho$. That is a consequence of the fact that relativistically, the contribution to the gravitational energy comes from $\rho_{Komar} = \rho + 3p$ \cite{TP} (for the perfect fluid) or $\rho_{Komar} = \rho + p_{r} + 2p_{\theta} = 2p_{\theta}$ (fo our imperfect fluid) \cite{HC5}. We could check that by calculating the Tolman-Komar energy \cite{TP}
\begin{equation}
W = 2 \int(T_{ab} - \frac{1}{2} g_{ab}T^{c}_{~c})u^{a} u^{b} N\sqrt{h} d^{3}x ,
\label{2.8}
\end{equation}
where $u^{a}$ is given by (2.2), $N = \sqrt{-g_{tt}}$ is the lapse function and $h$ is the determinant of the spatial 3 - metric, $h_{ab} = g_{ab} + u_{a} u_{b}$. Eq.(2.8) yields
\begin{equation}
W = 4\pi \int_{0}^{\infty}(\rho + p_{r} + 2p_{\theta}) N r^{2}\sqrt{h}~ dr = -\frac{b}{2}.
\label{2.9}
\end{equation}
 One observes in (2.9) the decisive contribution to W from the negative pressure. It is not surprising that $W<0$; it is a consequence of the repulsive character of the forces created by the source of curvature. A negative W one obtains also for the de Sitter spacetime in static coordinates, where Komar energy is proportional to the radius of curvature.

\section{Star interior geometry}
Let us apply the above recipe inside a relativistic star. We mean the star is filled with an imperfect fluid having negative pressures, up to a radius R, with $r\leq R$, so that the fluid is confined to a sphere of radius R and the metric inside is given by (2.1). In all the previous equations we have to replace the limit to infinity with $r \rightarrow R$. For example, the Komar energy will not be given by (2.9) but $W(R,b) = -\frac{b}{2}e^{-\frac{b}{R}}$. 

As the exterior spacetime should be Schwarzschild, the Darmois-Israel junction conditions have to be satisfied at $r = R$. From the first junction condition one finds that the relation $1 - e^{-\frac{b}{R}} = 1 - \frac{2M}{R}$ would be obeyed. Hence,
\begin{equation}
b = R ~ln\frac{R}{2M},
\label{3.1}
\end{equation}
where $M$ is the Schwarzschild mass ($R>2M$ is mandatory for to obtain $b>0$). Using (3.1), the Komar energy becomes $W = -Mln\frac{R}{2M}$, with $W = -M$ if $R = 2eM$. In addition, when, say, $R = e^{10}\cdot 2M, W = -10 M$ and if $R = 2M + 10^{-4}\cdot 2M, W \approx -10^{-4}M$. In other words, when $R\rightarrow 2M$ from above, the Komar energy tends to zero, and if $R>>2M$, $W$ is negative and very large, i.e., $|W|>>M$.

We notice that the constant $b$ depends on two parameters, R and M. For instance, an R very close to $2M$ gives a very small $b$ w.r.t. R. Take $R = 2M + \epsilon$ ($\epsilon $ small) and, say, $r>>b$ leads to $r>>R~ln(1 + \frac{\epsilon}{2M}) \approx \frac{R}{2M}\epsilon$. If $\epsilon = 2M\cdot 10^{-3}$, we get $r>>10^{-3}R$, a possible situation. In other words, all situations are possible, i.e., $r\geq b$ or $r<b$ (when $R>2eM$). To get quasi-flat space inside, the condition $R>>2M$ should be observed. Anyway, when the star, viewed frm outside, tends to become a black hole ($R \rightarrow 2M$), $b/R$ becomes very small and $1 - e^{-\frac{b}{R}} \approx 0$, such that $r = R$ becomes a horizon even for an inner observer. That can be also seen from the new form of the interior metric
	 \begin{equation}
  ds^{2} = -\left[1- \left(\frac{2M}{R}\right)^{\frac{R}{r}}\right] ~dt^{2} + \frac {dr^{2}}{1- \left(\frac{2M}{R}\right)^{\frac{R}{r}}} + r^{2} d \Omega^{2}
 \label{3.2}
 \end{equation}
We wish now to compare our geometry (3.2) with the Schwarzschild interior geometry \cite{MM2}, which is known to be conformally-flat. It is infered by supposing that the energy density is constant, a hypothesis presumed to be unphysical \cite{MM2}. Moreover, the pressure and the scalar curvature are divergent at the horizon $r_{H} = 3R\sqrt{1 - \frac{4R}{9M}}$. In addition, the central pressure becomes divergent when star radius reaches Buchdahl value $9M/4$. 

In contrast, the spacetime (3.2) is free of singularities (no infinite curvature) and $\rho$ and the pressures are finite everywhere. However, the weak, null and dominant energy conditions are satisfied only in the region $r\geq b(\sqrt{5} - 1)/2$, which is valid for $R\leq 10M$, namely when the star is on the verge to become a black hole. Nevertheless, the strong energy condition is not obeyed because $\rho + p_{r} + 2p_{\theta} <0$ everywhere.

As an example, we apply the previous results to the case of a neutron star (NS). We take the radius of the NS to be $R = 15 Km$, its mass of the order of Solar mass. With $2\cdot M_{\odot} \approx 3 Km$, we get $b = Rln5>R$ and $r_{max} \approx 8.76 Km<R$, so that $\rho_{max} = 1.36\cdot 10^{15} g/cm^{3}$, a value very close to the known densities of a neutron star. If we evaluate $\rho$ at, say,  $r = R/2 = 7.5 Km <r_{max}$, we get $\rho (R/2) = 2.2\cdot 10^{14} g/cm^{3}$, one sixth from the value of $\rho_{max}$.

Let us turn now to the problem of matching conditions at the interface. For to write the second junctions conditions, we need the extrinsic curvature $K_{ab}$ at the boundary $r = R$. Having spherical symmetry and static conditions, we make use of the expressions \cite{KKP}
\begin{equation}
	K_{ab} = -\frac{f'}{\sqrt{f}} u_{a} u_{b} + \frac{\sqrt{f}}{r} q_{ab},
 \label{3.3}
 \end{equation}
whence
\begin{equation}
	K = \gamma^{ab} K_{ab} = \frac{f'}{2\sqrt{f}} + 2\frac{\sqrt{f}}{r},
 \label{3.4}
 \end{equation}
where $f = -g_{tt}, u_{a} = (\sqrt{f}, 0, 0, 0)$, $\gamma_{ab} = g_{ab} -  n_{a} n_{b}$ is the induced metric on $r = const.$ surface, with $n^{a} = (0, \sqrt{f}, 0, 0)$ its normal vector and $q_{ab} = \gamma_{ab} + u_{a} u_{b}$ is the induced metric on a two-surface of constant $t$ and $r$. One obtains, in terms of the coordinates inside (-) the star
\begin{equation}
K^{-}_{tt} = \frac{b}{2r^{2}} \sqrt{f} e^{-\frac{b}{r}},~~~K^{-}_{\theta \theta} = r \sqrt{f},~~~K^{-} = -\frac{b}{2r^{2}\sqrt{f}} e^{-\frac{b}{r}} + \frac{2}{r} \sqrt{f},
 \label{3.5}
 \end{equation}
where all quantities are evaluated at $r = R$. For the Schwarzschild exterior (+) region we have \cite{HC4}
\begin{equation}
K^{+}_{tt} = -\frac{M}{r^{2}} \sqrt{1 - \frac{2M}{r}} ,~~~K^{+}_{\theta \theta} = r \sqrt{1 - \frac{2M}{r}},~~~K^{+} = \frac{2r - 3M}{r^{2}\sqrt{1 - \frac{2M}{r}}} ,
 \label{3.6}
 \end{equation}
evaluated at $r = R$. The discontinuity of the extrinsic curvature $[K_{ab}] = K_{ab}^{+} - K_{ab}^{-}$ is related to the stress tensor $S_{ab}$ on the hypersurface $r = R$ through the Lanczos equation
\begin{equation}
[K_{ab}] - \gamma_{ab}[K] = -8\pi S_{ab}
 \label{3.7}
 \end{equation}
Let us suppose $S_{ab}$ to have the perfect fluid form \cite{CL}
\begin{equation}
S_{ab} = (\rho_{s} + p_{s}) u_{a} u_{b} + p_{s}\gamma_{ab},
 \label{3.8}
 \end{equation}
	where $\rho_{s}$ stands for the surface energy density, $p_{s}$ is the surface pressure, with $\sigma = -p_{s}$ the surface tension. By means of the expression of $b$ from (3.1) we get
	\begin{equation}
	[K_{tt}] - \gamma_{tt}[K] = \frac{2R - 4M}{R^{2}} \sqrt{1 - \frac{2M}{R}} - \frac{2}{R} \left(1 - \frac{2M}{R}\right)^{3/2} = 0.
 \label{3.9}
 \end{equation}
Therefore, we have $S_{tt} = 0$ which yields $\rho_{s} = 0$. Now, from the $\theta \theta$-component of (3.8), one finds that 
		\begin{equation}
		M \left(1 + \frac{b}{R}\right) = -8\pi R^{2} p_{s} \sqrt{1 - \frac{2M}{R}}, 
 \label{3.10}
 \end{equation}
 which shows that the surface tension $\sigma = -p_{s}$ is positive. We may express $\sigma$ in terms of the difference of the normal accelerations $a^{b}n_{b}$ for a static observer sitting at $r = R$ 
		\begin{equation}
	(a^{b}n_{b})_{+} - (a^{b}n_{b})_{-}	= \frac{M}{R^{2}\sqrt{1 - \frac{2M}{R}}} + \frac{b e^{-\frac{b}{R}}}{2R^{2} \sqrt{1 - \frac{2M}{R}}}  = \frac{M \left(1 + \frac{b}{R}\right)}{R^{2}\sqrt{1 - \frac{2M}{R}}} ,
 \label{3.11}
 \end{equation}
which gives the same expression for $\sigma$ as (3.10) (see also \cite{MM2}, where the authors worked instead with surface gravities and got the same result). Even though the surface tension is cohesive ($\sigma>0$) the repulsive gravity in the interior creates the possibility to have a balance of forces required for establishing a hydrostatic equilibrium \cite{CK}.
 
Our aim now is to compare the above expression of $\sigma$ with that given by the Young-Laplace equation \cite{HC4}
		\begin{equation}
		\rho = -p_{r} = \frac{2\sigma}{r},
 \label{3.12}
 \end{equation}
with $\rho$ and $p_{r}$ given by (2.6). From (3.12) one obtains
\begin{equation}
	\sigma = \frac{M \left(1 + \frac{b}{R}\right)}{8\pi R^{2}}
 \label{3.13}
 \end{equation}
which is not equal to the expression of $\sigma$ obtained from (3.10), because of the relativistic factor $\sqrt{1 - \frac{2M}{R}}$. The origin of this mismatch is due, in our view, to the classical nature of (3.11).

\section{PG coordinates. Geodesics}
We follow the standard procedure to pass to the Painleve-Gullstrand coordinates, from the original Schwarzschild-form (2.1) of the geometry supposed to be valid inside the star and introduce a new time coordinate 
\begin{equation}
T = t - g(r).
 \label{4.1}
 \end{equation}
When we put (4.1) in (2.1) and choose the function $g(r)$ such that
\begin{equation}
g'(r) = \frac{1}{f(r)} e^{-\frac{b}{2r}},
 \label{4.2}
 \end{equation}
we get $g_{rr} = 1$ and therefore, the metric (2.1) acquires the PG form
\begin{equation}
ds^{2} = -(1 - e^{-\frac{b}{r}}) dT^{2} - 2~ e^{-\frac{b}{2r}} dTdr +dr^{2} + r^{2} d \Omega^{2},
 \label{4.3}
 \end{equation}
which, at $r = R$ matches the standard Schwarzschild line-element in PG coordinates. The geometry (4.3) is an exact solution of gravitational equations when the source is the same energy-momentum tensor $T_{~a}^{b}$ from (2.5), with the same expressions for the energy density and pressures due to the fact that the metric is static and the transformation (4.1) changes only the time variable.

Let us take the following congruence of observers with the velocity vector field
\begin{equation}
u^{a} = (1, e^{-\frac{b}{2r}}, 0, 0),~~~u^{a}u_{a} = -1.
 \label{4.4}
 \end{equation}
One easily finds that the form (4.4) of the velocity field gives us $a^{b}= 0$, such that $u^{a}$ is tangent to the timelike geodesics, comoving with the fluid. The normal vector orthogonal to $u^{a}$ is given by $n_{a} = (-e^{-\frac{b}{2r}}, 1, 0, 0)$. From the general expression (2.5) of the stress tensor one obtains again zero energy flux density, namely $q^{a} = 0$. In contrast, we get  a nonzero shear tensor and expansion scalar in the PG coordinates, inside the star
\begin{equation}
\sigma^{r}_{~t} = -\sigma^{r}_{~r} = 2\sigma^{\theta}_{~\theta} = 2\sigma^{\phi}_{~\phi} = \frac{2}{3r}\left(1 - \frac{b}{2r}\right) e^{-\frac{b}{r}}
 \label{4.5}
 \end{equation}
with $\sigma^{a}_{~a} = 0$ and $\Theta = \frac{2}{r}(1 + \frac{b}{4r}) e^{-\frac{b}{2r}}$.\\ 

\textbf{Timelike radial geodesics}\\
From the expression (4.4) of $u^{a}$ we observe that $T$ represents the proper time and therefore 
\begin{equation}
v \equiv u^{r} = \frac{dr}{dT} = e^{-\frac{b}{2r}}.
 \label{4.6}
 \end{equation}
However, the trajectories $r(T)$ cannot be determined exactly but from (4.6) we may estimate that $v \rightarrow 0$ at the origin and $v(R) = e^{-\frac{b}{2R}} = \sqrt{\frac{2M}{R}}$, the Newtonian escape velocity. Consequently, $0<v\leq \sqrt{2M/R}$ inside the star. It is worth noting that $v_{max} = \sqrt{2M/R}$ matches exactly the escape velocity on the surface, measured from outside the star, using the exterior PG coordinates. If we took $g_{rT} >0$ in the line-element (4.3), the timelike geodesics would become ingoing, with $v = -e^{-\frac{b}{2r}}$ and $-\sqrt{2M/R} \leq v <0$. \\

\textbf{Null radial geodesics}\\
The null radial geodesics are directly obtained from (4.3), with $ds^{2} = 0$
\begin{equation}
 -(1 - e^{-\frac{b}{r}})  - 2~ e^{-\frac{b}{2r}} \frac{dr}{dT} + \left(\frac{dr}{dT}\right)^{2} = 0
 \label{4.7}
 \end{equation}
which yields $\frac{dr}{dT} = e^{-\frac{b}{2r}} \pm{1}$. Only the minus sign is convenient and we have
\begin{equation}
  V \equiv \frac{dr}{dT} = e^{-\frac{b}{2r}} - 1,
 \label{4.8}
 \end{equation}
which are ingoing null geodesics, with $-1 \leq V < \sqrt{2M/R} - 1$. In contrast, $g_{rT} >0$ will lead us to 
\begin{equation}
  V = 1 - e^{-\frac{b}{2r}} >0,
 \label{4.9}
 \end{equation}
(outgoing geodesics), whence one finds that $1>V>1 - \sqrt{2M/R}$. We notice also that the scalar expansion $\Theta$ is positive when the timelike geodesics are outward and it changes sign if $g_{rT} >0$ (ingoing timelike geodesics).

\section{Conclusions}
It is a known fact that the interior Schwarzschild solution discovered by Schwarzschild in 1916 has a lot of drawbacks, such as divergent pressure and scalar curvature at the horizon. In addition, the assumption of constant density is presumed unphysical. 

We looked in this paper for a regular star interior, with nonsingular energy density and pressures, by means of an exponential regulator. Because of the negative pressures the gravitational field inside is repulsive, as for a de Sitter geometry in static coordinates. As a consequence, the Komar energy is negative. With the help of the matching conditions at the star surface we found the constant length $b$ from the metric and the surface tension $\sigma$ of the boundary with the exterior Schwarzschild metric. The timelike and null geodesics are studied in Painleve-Gullstrand coordinates.

\end{document}